\documentclass[twocolumn,prl,showpacs,showkeys,amsmath,amssymb]{revtex4}
\usepackage{amsthm,url}

\theoremstyle{plain}

\begin{document}
\title{Insecurity Of Imperfect Quantum Bit Seal}
\author{H.~F. Chau}
\email{hfchau@hkusua.hku.hk}
\affiliation{Department of Physics, University of Hong Kong, Pokfulam Road,
 Hong Kong}
\affiliation{Center of Theoretical and Computational Physics, University of
 Hong Kong, Pokfulam Road, Hong Kong}
\date{\today}

\begin{abstract}
 Quantum bit seal is a way to encode a classical bit quantum mechanically so
 that everyone can obtain non-zero information on the value of the bit.
 Moreover, such an attempt should have a high chance of being detected by an
 authorized verifier. Surely, a reader looks for a way to get the maximum
 amount of information on the sealed bit and at the same time to minimize her
 chance of being caught. And a verifier picks a sealing scheme that maximizes
 his chance of detecting any measurement of the sealed bit. Here, I report a
 strategy that passes all measurement detection procedures at least half of the
 time for all quantum bit sealing schemes. This strategy also minimizes a
 reader's chance of being caught under a certain scheme. In this way, I extend
 the result of Bechmann-Pasquinucci \emph{et al.} by proving that quantum seal
 is insecure in the case of imperfect sealed bit recovery.
\end{abstract}

\pacs{03.67.Dd, 03.67.Hk, 89.20.Ff, 89.70.+c}
\keywords{Information disturbance tradeoff, Post-modern quantum cryptography,
 Quantum seal}
\maketitle
\emph{Introduction ---}
 We sometimes put an important document, such as a will, in an envelop with
 sealed wax so that others can open it only by breaking the wax. The intactness
 of the wax seal, therefore, shows that the document has not been opened. It is
 useful to extend the concept of physical wax seal to the digital world. Yet,
 no classical digital sealing scheme is unconditionally secure as one can, in
 principle, copy all the bits without being caught.

 Recently, Bechmann-Pasquinucci explored the possibility of sealing a classical
 bit quantum mechanically. In her scheme, each bit of classical message is
 encoded as three qubits out of which one of them is erroneous. Using single
 qubit measurement along the standard basis plus the classical $[3,1,3]_2$
 majority vote code, everyone can obtain the original classical bit with
 certainty. Moreover, an authorized verifier, who knows some extra information
 on the erroneous qubit, may check if someone has extracted the encoded
 classical bit with non-negligible probability \cite{quantseal}. Following a
 similar line, Chau proposed quantum bit seal based on quantum error correcting
 code that applied to quantum messages \cite{oldversion}. Recently, Singh and
 Srikanth \cite{anotherseal} as well as He \cite{anotherseal2,stringseal}
 separately constructed quantum seals by extending the Bechmann-Pasquinucci
 protocol in different ways. In particular, He constructed a quantum bit string
 seal \cite{stringseal}.

 The above five quantum sealing schemes can be divided into two types. The
 first two are perfect quantum seals in the sense that a reader can obtain the
 classical message with certainty. The last three are imperfect for a reader
 cannot do so accurately.

 The density matrices used to encode any two distinct classical messages in a
 perfect quantum seal must be orthogonal. Hence, one may devise a collective
 measurement to read out the sealed message without disturbing the encoded
 state. (A quantum computer may be required to perform the collective
 measurement though.) This is precisely the idea used by Bechmann-Pasquinucci
 \emph{et al.} to prove the insecurity of all perfect quantum seals
 \cite{impossibleperfect}. However, their proof does not apply to imperfect
 seals. Therefore, it is instructive to study the security of imperfect seals.
 Recently, He analyzed the security of imperfect quantum bit seals and proved
 certain bounds regarding the information gain and the measurement detection
 probability \cite{bound}. However, his bound is not tight.

 In this Letter, I study the relation between the information on the sealed bit
 obtained by a reader and the probability of measurement detection by a
 verifier. In particular, I find a cheating strategy that obtains non-zero
 amount of information on the sealed bit but escapes detection at least half of
 the time for all quantum bit sealing schemes. Moreover, this strategy is
 optimal for a certain bit sealing scheme in the sense that no strategy with
 the same power of distinguishing the two sealed bits has a higher chance of
 avoiding measurement detection under that scheme. (Implementing this cheating
 strategy may require a quantum computer.) Consequently, I conclude that all
 imperfect quantum bit seals are insecure.
 
\par\medskip
\emph{Quantum Bit Seal ---}
 A quantum bit seal is a method for Alice to encode a bit in such a way that
 any member of the public, say Bob, can recover the original bit with
 probability greater than a half. Moreover, any recovery process must disturb
 the encoded state so that the intactness of the state plays the role of a wax
 seal. By checking its intactness, an authorized verifier may correctly detect
 any measurement of the value of the sealed bit with a certain non-zero
 probability. Without loss of generality, one may assume that the encoded state
 $i$ Alice prepared is a pure state given by
\begin{equation}
 |\tilde{\psi}_i\rangle = \sum_j \lambda_{ij} |\psi_{ij}\rangle_B \otimes
 |\phi_j\rangle_A \label{E:Def_psi}
\end{equation}
 for $i=0,1$, where $|\psi_{ij}\rangle$'s are normalized states that are not
 necessarily mutually orthogonal, and $|\phi_j\rangle$'s are orthonormal
 states. (If the states prepared by Alice are mixed, she can always purify
 them. More importantly, the use of pure states increases the verifier's chance
 to detect a cheating Bob.) She makes the particles labeled by the subscript
 ``B'' publicly accessible, and keeps those labeled by the subscript ``A'' for
 authorized verifiers only. So, from Bob's point of view, the sealed state is
 either $\rho_0$ or $\rho_1$, where
\begin{equation}
 \rho_i = \sum_j |\lambda_{ij} |^2 |\psi_{ij}\rangle\langle\psi_{ij}| ~.
 \label{E:Def_rho}
\end{equation}

 The maximum achievable classical $L_1$ distance between probability
 distributions arising from measurements performed on $\rho_0$ and $\rho_1$ is
 equal to the trace distance between the two density matrices \cite{NC}
\begin{equation}
 q_\text{max} = D(\rho_0,\rho_1) \equiv \frac{1}{2} \mbox{Tr} |\rho_0 - \rho_1
 | ~. \label{E:Def_trace_distance}
\end{equation}
 The higher the value of $q_\text{max}$, the higher the chance for Bob to
 correctly extract the sealed bit. In fact, a quantum bit seal is perfect if
 $q_\text{max} = 1$; and it is imperfect if $0\leq q_\text{max} < 1$.
 (Actually, the case of $q_\text{max} = 0$ is not a quantum bit seal at all for
 it gives no information on the original state.)

 To gain information on the sealed state, Bob performs a positive
 operator-valued measure (POVM) measurement ${\mathcal E}$ on the publicly
 accessible state. The classical $L_1$ distance between probability
 distributions arising from the measurement ${\mathcal E}$ on $\rho_0$ and
 $\rho_1$ equals $q\equiv D({\mathcal E}(\rho_0),{\mathcal E} (\rho_1))$. From
 Alice's point of view, the entire entangled state becomes ${\mathcal E}
 \otimes I (|\tilde{\psi}_i\rangle\langle\tilde{\psi}_i|) \equiv
 \tilde{\mathcal E} (|\tilde{\psi}_i\rangle\langle\tilde{\psi}_i|)$. Amongst
 all ${\mathcal E}$ whose measurement results on $\rho_0$ and $\rho_1$ have the
 same classical $L_1$ distance $q\in [0,q_\text{max}]$, Bob would like to pick
 the one that minimizes the chance of being detected by a verifier. Since it is
 equally likely for the sealed bit to be $0$ or $1$ and the probability of
 catching the reader Bob is at most equal to the fidelity of the resultant
 state as $|\tilde{\psi}_i\rangle$, Bob's aim is equivalent to finding a
 quantum operation $\tilde{\mathcal E} = {\mathcal E} \otimes I$ that maximizes
 the average fidelity
\begin{equation}
 \bar{F} = \frac{1}{2} \sum_{i=0}^1 \langle\tilde{\psi}_i | \tilde{\mathcal E}
 (|\tilde{\psi}_i\rangle\langle\tilde{\psi}_i|) | \tilde{\psi}_i\rangle \equiv
 \left< \mbox{Tr} [ \tilde{\mathcal E}(\tilde{\rho}_i) \,\tilde{\rho}_i ]
 \right>_i \label{E:Def_F} ~.
\end{equation}
 On the other hand, Alice has the freedom to pick the sealing scheme. Of
 course, she would like to choose the one that minimizes $\bar{F}$.
 Consequently, the average fidelity for the optimal cheating strategy
 $\min_\text{Alice} \max_\text{Bob} \bar{F}(q,q_\text{max})$ is found by first
 taking the maximum over all possible POVM measurements ${\mathcal E}$ used by
 Bob with $q = D({\mathcal E} \otimes I (\tilde{\rho}_0), {\mathcal E} \otimes
 I (\tilde{\rho}_1))$ for a given sealing scheme, and then by taking the
 minimum over all quantum bit seals with trace distance $q_\text{max}$ chosen
 by Alice. Here, I prove that
\begin{equation}
 \min_\text{Alice} \max_\text{Bob} \bar{F}(q,q_\text{max}) = \frac{1+
 q_\text{max}^2}{2} + \frac{1-q_\text{max}^2}{2} \left( 1-
 \frac{q^2}{q_\text{max}^2} \right)^\frac{1}{2} \label{E:opt_F}
\end{equation}
 for all $0 < q \leq q_\text{max}$ and that
\begin{equation}
 \min_\text{Alice} \max_\text{Bob} \bar{F}(0,q_\text{max}) = 1 ~.
 \label{E:opt_F_0}
\end{equation}

\par\medskip
\emph{The Optimal Cheating Strategy ---}
 Let me introduce a few notations before reporting the optimal cheating
 strategy that works for $q>0$. (Note that this strategy is optimal in the
 sense stated in the above paragraph. That is to say, this is the best strategy
 for Bob to work against the most stringent sealing scheme devised by Alice. It
 is possible, however, to find a strategy with a higher value of average
 fidelity if Alice uses a less stringent scheme.) I write $\rho_0 - \rho_1 =
 Q_0 - Q_1$, where $Q_i$'s are positive operators with orthogonal support.
 Suppose $\Pi_0$ satisfying $\Pi_0 Q_0 = Q_0$ is a projector whose support is
 orthogonal to that of $Q_1$, and let $\Pi_1 = I - \Pi_0$ be its complementary
 projector. (Clearly, these projectors exist and are non-zero if and only if
 $q_\text{max} > 0$. In addition, these non-zero projectors are uniquely
 determined on the support of $\rho_0 + \rho_1$ if and only if $\rho_0 -
 \rho_1$ and $\rho_0 + \rho_1$ have a common support.) Then, the classical
 $L_1$ distance between probability distributions arising from the POVM
 measurement with measurement operators $\{ \Pi_0, \Pi_1 \}$ on $\rho_0$ and
 $\rho_1$ equals $q_\text{max}$ \cite{NC,Helstrom}. By denoting
\begin{equation}
 a = \langle\tilde{\psi}_0|\Pi_0 \otimes I |\tilde{\psi}_0\rangle ~,
 \label{E:Def_a}
\end{equation}
 one has
\begin{eqnarray}
 \langle\tilde{\psi}_0|\Pi_1 \otimes I |\tilde{\psi}_0\rangle & = & 1-a ~,
  \label{E:conseq1} \\
 \langle\tilde{\psi}_1|\Pi_0 \otimes I |\tilde{\psi}_1\rangle & = & a-
  q_\text{max} \label{E:conseq2}
\end{eqnarray}
 and
\begin{equation}
 \langle\tilde{\psi}_1|\Pi_1 \otimes I |\tilde{\psi}_1\rangle = 1-a+
  q_\text{max} ~. \label{E:conseq3}
\end{equation}
 Clearly, $q_\text{max} \leq a \leq 1$.

 For any fixed $0 < q \leq q_\text{max}$, let
\begin{equation}
 M_0 = \left[ \frac{1}{2} \left( 1+\frac{q}{q_\text{max}} \right)
 \right]^\frac{1}{2} \Pi_0 + \left[ \frac{1}{2} \left( 1-
 \frac{q}{q_\text{max}} \right) \right]^\frac{1}{2} \Pi_1 \label{E:Def_M_0}
\end{equation}
 and
\begin{equation}
 M_1 = \left[ \frac{1}{2} \left( 1-\frac{q}{q_\text{max}} \right)
 \right]^\frac{1}{2} \Pi_0 + \left[ \frac{1}{2} \left( 1+
 \frac{q}{q_\text{max}} \right) \right]^\frac{1}{2} \Pi_1 ~. \label{E:Def_M_1}
\end{equation}
 Then $M_0^\dag M_0 + M_1^\dag M_1 = I$. So, the operation
 ${\mathcal E}_{q,q_\text{max}}$ given by
\begin{equation}
 {\mathcal E}_{q,q_\text{max}} (\rho_i) = M_0 \rho_i M_0^\dag + M_1 \rho_i
 M_1^\dag \label{E:Def_opt_POVM}
\end{equation}
 is a POVM measurement with $D({\mathcal E}_{q,q_\text{max}}(\rho_0),
 {\mathcal E}_{q,q_\text{max}}(\rho_1))$ $= q$. Bob applies this POVM
 measurement to the publicly accessible quantum particles and deduces the value
 of the sealed bit as $i$ if the measurement outcome is $i$. (Bob may require a
 quantum computer in order to implement this measurement.)

 From Eqs.~(\ref{E:Def_a})--(\ref{E:Def_M_1}), the probability of correctly
 determine the sealed bit equals
\begin{equation}
 \mbox{Pr} = \left< \langle\tilde{\psi}_i|M_i^\dag M_i|\tilde{\psi}_i\rangle
 \right>_i = \frac{1+q}{2} ~. \label{E:correct_pr}
\end{equation}

 Similarly, the average fidelity of the resultant state caused by this cheating
 strategy is given by
\begin{eqnarray}
 \bar{F}(q) & = & \frac{1}{2} \sum_{i,j=0}^1 \left| \langle\tilde{\psi}_i|M_j
  \otimes I |\tilde{\psi}_i\rangle \right|^2 \nonumber \\
 & = & \frac{1}{2} \sum_{i=0}^1 \left[ \sum_{j=0}^1 \mbox{Tr} \left( \Pi_j
  \otimes I \tilde{\rho}_i \Pi_j \otimes I \tilde{\rho}_i \right) \right.
  \nonumber \\
 & & ~+ \left. 2 \left( 1-\frac{q^2}{q_\text{max}^2} \right)^{1/2}
  \mbox{Tr} \left( \Pi_0 \otimes I \tilde{\rho}_i \Pi_1 \otimes I
  \tilde{\rho}_i \right) \right] \nonumber \\
 & = & (1-2a)(1+q_\text{max}) + 2a^2 + q_\text{max}^2 - [2a^2 \nonumber \\
 & & ~+ (q_\text{max}-2a)(1+q_\text{max})] \left( 1-\frac{q^2}{q_\text{max}^2}
  \right)^\frac{1}{2} ~. \label{E:fbar}
\end{eqnarray}
 Note that $\bar{F}(q)$ attains its minimum value when $a=(1+q_\text{max})/2$.
 More importantly, from the construction of $\Pi_i$'s, Alice may force $a$ to
 take on the value $(1+q_\text{max})/2$ by choosing $|\tilde{\psi}_i\rangle$'s
 in such a way that the supports of $\rho_0 - \rho_1$ and $\rho_0 + \rho_1$
 agree and that $\langle\tilde{\psi}_0|\Pi_0|\tilde{\psi}_0\rangle = (1+
 q_\text{max})/2$. (Such $|\tilde{\psi}_i\rangle$'s certainly exist. One
 possible choice is $|\tilde{\psi}_i\rangle = \{ [1+(-1)^i q_\text{max}]/2
 \}^{1/2} |0\rangle_B + \{ [1-(-1)^i q_\text{max}]/2 \}^{1/2} |1\rangle_B$ for
 $i=0,1$.) Once Alice has picked such $|\tilde{\psi}_i\rangle$'s, the average
 fidelity of the resultant state after applying the POVM measurement
 ${\mathcal E}_{q,q_\text{max}}$ is
\begin{equation}
 \bar{F}(q) = \frac{1+q_\text{max}^2}{2} + \frac{1-q_\text{max}^2}{2} \left( 1
 - \frac{q^2}{q_\text{max}^2} \right)^\frac{1}{2} ~. \label{E:opt_fbar}
\end{equation}
 This $\bar{F}(q)$ sets the lower bound for the value of
 $\min_\text{Alice} \max_\text{Bob} \bar{F}(q,q_\text{max})$ in the case of
 $q>0$.

\par\medskip
\emph{The Proof Of Optimality ---}
 I prove the optimality of the cheating strategy ${\mathcal E}_{q,
 q_\text{max}}$ by explicitly constructing a quantum bit seal whose value of
 $\bar{F}(q)$ is upper bounded by the right hand side of Eq.~(\ref{E:opt_fbar})
 for any cheating strategy used by Bob. I claim that
\begin{eqnarray}
 |\tilde{\psi}_i\rangle & = & \frac{\sqrt{1-q_\text{max}}}{2} \{ [ |0\rangle_B
  + (-1)^i |1\rangle_B] \otimes |0\rangle_A \nonumber \\
 & & ~+ [|0\rangle_B - (-1)^i |1\rangle_B] \otimes |1\rangle_A \} \nonumber \\
 & & ~+ q_\text{max}^{1/2} |i\rangle_B \otimes |2\rangle_A \label{E:opt_scheme}
\end{eqnarray}
 for $i=0,1$ is such a scheme.

 Clearly, $\rho_i = \mbox{Tr}_A (|\tilde{\psi}_i\rangle\langle\tilde{\psi}_i|)
 = \mbox{diag} ( [1+(-1)^i q_\text{max}]/2,$ $[1-(-1)^i q_\text{max}]/2 )$.
 Thus, $D(\rho_0,\rho_1) = q_\text{max}$.

 Recall that the most general measurement strategy for Bob is to perform a POVM
 measurement with measurement operators $N_j$'s, where
\begin{equation}
 N_j = \left[ \begin{array}{cc}
  \alpha_j & \beta_j \\
  \gamma_j & \delta_j
 \end{array} \right] ~. \label{E:Def_N_j}
\end{equation}
 To qualify as a POVM measurement whose measurement results on $\rho_0$ and
 $\rho_1$ give probability distributions with classical $L_1$ distance $q$, one
 requires
\begin{equation}
 \sum_j ( |\alpha_j|^2 + |\gamma_j|^2 ) = 1 = \sum_j ( |\beta_j|^2 +
 |\delta_j|^2 ) ~, \label{E:const1}
\end{equation}
\begin{equation}
 \sum_j ( \alpha_j \bar{\beta}_j + \gamma_j \bar{\delta}_j ) = 0
 \label{E:const2}
\end{equation}
and
\begin{equation}
 q_\text{max} \sum_j \left| |\alpha_j|^2 + |\gamma_j|^2 - |\beta_j|^2 -
 |\delta_j|^2 \right|  = 2q ~. \label{E:const3}
\end{equation}

 The average fidelity of the state after applying ${\mathcal E}_{q,
 q_\text{max}}$ equals
\begin{eqnarray}
 \bar{F}(q) & = & \frac{1+q_\text{max}^2}{4} \sum_j ( |\alpha_j|^2 +
  |\delta_j|^2 ) \nonumber \\
 & & ~+ \frac{1-q_\text{max}^2}{4} \sum_j (\alpha_j \bar{\delta}_j +
  \bar{\alpha}_j \delta_j) ~. \label{E:F_stringent}
\end{eqnarray}
 The constraint in Eq.~(\ref{E:const1}) implies that
\begin{equation}
 \sum_j \left( |\alpha_j|^2 + |\delta_j|^2 \right) \leq 2 ~.
 \label{E:inequality1}
\end{equation}
 And by constrained maximization, one concludes that
\begin{equation}
 \sum_j (\alpha_j \bar{\delta}_j + \bar{\alpha}_j \delta_j) \leq 2 \left( 1 -
 \frac{q^2}{q_\text{max}^2} \right)^\frac{1}{2} ~. \label{E:inequality2}
\end{equation}
 Consequently, $\bar{F}(q)$ for this scheme is less than or equal to the right
 hand side of Eq.~(\ref{E:opt_fbar}). Therefore, the optimality of
 ${\mathcal E}_{q,q_\text{max}}$ together with the validity of
 Eq.~(\ref{E:opt_F}) are proven.

 It remains to consider the case of $q=0$. In this case, Bob simply applies the
 identity operator to the state. Hence, $\min_\text{Alice} \max_\text{Bob}
 \bar{F}(0,q_\text{max}) = 1$. Note that the positive operators $Q_0$ and $Q_1$
 are non-zero if and only if $q_\text{max} > 0$. The sudden change in the
 dimensions of $Q_i$'s is probably the reason why $\min_\text{Alice}
 \max_\text{Bob} \bar{F}(q,q_\text{max})$ is discontinuous at $(0,0)$.

\par\medskip
\emph{Discussions ---}
 In short, I have studied the relation between the information gain on the
 sealed qubit and the probability of sealed bit measurement detection. In
 particular, I find a cheating strategy ${\mathcal E}_{q,q_\text{max}}$ whose
 probability of passing any measurement detection procedure is greater than or
 equal to the right hand side of Eq.~(\ref{E:opt_fbar}). Since this probability
 is at least $1/2$, I conclude that both perfect and imperfect quantum bit
 seals are insecure. This strategy reduces to the one used by
 Bechmann-Pasquinucci \emph{et al.} \cite{impossibleperfect} in the case of a
 perfect seal. More importantly, the cheating strategy ${\mathcal E}_{q,
 q_\text{max}}$ is optimal in the sense that it has the least possible chance
 of being detected by a verifier for the most stringent quantum bit sealing
 scheme. Although all quantum states presented in this Letter live in finite
 dimensional Hilbert spaces, the arguments used in the proof are completely
 general and are applicable also to states living in infinite dimensional
 Hilbert spaces.

 At least two lines of followup researches are worth to conduct. One line is to
 investigate the security of imperfect quantum bit string seal. Although one
 may apply the cheating strategy reported above to correctly read out a single
 bit in any quantum bit string seal with a small chance of being caught, the
 possibility (or impossibility) of correctly determining the values of a large
 proportion of bits without being caught remains unclear. The other line of
 research is to look for a quantum bit sealing scheme whose security is based
 on computational (such as the inefficiency in computing $Q_0$ and $Q_1$) or
 hardware (such as the hardness of building a quantum computer) assumptions.

 Lastly, I remark that there is an irreparable gap in the ``security proof''
 of the quantum seal in my earlier manuscript in Ref.~\cite{oldversion} ---
 there is no way for Alice to force Bob to measure the publicly accessible
 qubits individually.

\begin{acknowledgments}
\par\medskip
\emph{Acknowledgments ---}
 This work is supported by the RGC grant HKU~7010/04P of the HKSAR Government.
\end{acknowledgments}

\bibliography{qc35.3}

\begin{thebibliography}{9}
\expandafter\ifx\csname natexlab\endcsname\relax\def\natexlab#1{#1}\fi
\expandafter\ifx\csname bibnamefont\endcsname\relax
  \def\bibnamefont#1{#1}\fi
\expandafter\ifx\csname bibfnamefont\endcsname\relax
  \def\bibfnamefont#1{#1}\fi
\expandafter\ifx\csname citenamefont\endcsname\relax
  \def\citenamefont#1{#1}\fi
\expandafter\ifx\csname url\endcsname\relax
  \def\url#1{\texttt{#1}}\fi
\expandafter\ifx\csname urlprefix\endcsname\relax\def\urlprefix{URL }\fi
\providecommand{\bibinfo}[2]{#2}
\providecommand{\eprint}[2][]{\url{#2}}

\bibitem[{\citenamefont{Bechmann-Pasquinucci}(2003)}]{quantseal}
\bibinfo{author}{\bibfnamefont{H.}~\bibnamefont{Bechmann-Pasquinucci}},
  \bibinfo{journal}{Int. J. Quant. Inform.} \textbf{\bibinfo{volume}{1}},
  \bibinfo{pages}{217} (\bibinfo{year}{2003}).

\bibitem[{\citenamefont{Chau}()}]{oldversion}
\bibinfo{author}{\bibfnamefont{H.~F.} \bibnamefont{Chau}},
  \emph{\bibinfo{title}{Sealing quantum message by quantum code}},
  \bibinfo{note}{quant-ph/0308146}.

\bibitem[{\citenamefont{Singh and Srikanth}(2005)}]{anotherseal}
\bibinfo{author}{\bibfnamefont{S.~K.} \bibnamefont{Singh}} \bibnamefont{and}
  \bibinfo{author}{\bibfnamefont{R.}~\bibnamefont{Srikanth}},
  \bibinfo{journal}{Physica Scripta} \textbf{\bibinfo{volume}{71}},
  \bibinfo{pages}{433} (\bibinfo{year}{2005}).

\bibitem[{\citenamefont{He}({\natexlab{a}})}]{anotherseal2}
\bibinfo{author}{\bibfnamefont{G.-P.} \bibnamefont{He}},
  \emph{\bibinfo{title}{Quantum secret sharing, hiding and sealing of classical
  data against collective measurement}},
  \bibinfo{howpublished}{quant-ph/0502091v1}.

\bibitem[{\citenamefont{He}({\natexlab{b}})}]{stringseal}
\bibinfo{author}{\bibfnamefont{G.-P.} \bibnamefont{He}},
  \emph{\bibinfo{title}{Quantum bit string sealing}},
  \bibinfo{howpublished}{quant-ph/0502091v3}.

\bibitem[{\citenamefont{Bechmann-Pasquinucci
  et~al.}(2005)\citenamefont{Bechmann-Pasquinucci, D'Ariano, and
  Macchiavello}}]{impossibleperfect}
\bibinfo{author}{\bibfnamefont{H.}~\bibnamefont{Bechmann-Pasquinucci}},
  \bibinfo{author}{\bibfnamefont{G.~M.} \bibnamefont{D'Ariano}},
  \bibnamefont{and}
  \bibinfo{author}{\bibfnamefont{C.}~\bibnamefont{Macchiavello}},
  \bibinfo{journal}{Int. J. Quant. Inform.} \textbf{\bibinfo{volume}{3}},
  \bibinfo{pages}{435} (\bibinfo{year}{2005}).

\bibitem[{\citenamefont{He}(2005)}]{bound}
\bibinfo{author}{\bibfnamefont{G.-P.} \bibnamefont{He}},
  \bibinfo{journal}{Phys. Rev. A} \textbf{\bibinfo{volume}{71}},
  \bibinfo{pages}{054304} (\bibinfo{year}{2005}).

\bibitem[{\citenamefont{Nielsen and Chuang}(2000)}]{NC}
\bibinfo{author}{\bibfnamefont{M.~A.} \bibnamefont{Nielsen}} \bibnamefont{and}
  \bibinfo{author}{\bibfnamefont{I.~L.} \bibnamefont{Chuang}},
  \emph{\bibinfo{title}{Quantum Computation and Quantum Information}}
  (\bibinfo{publisher}{CUP}, \bibinfo{address}{Cambridge},
  \bibinfo{year}{2000}), chap.~\bibinfo{chapter}{9}.

\bibitem[{\citenamefont{Helstrom}(1976)}]{Helstrom}
\bibinfo{author}{\bibfnamefont{C.~W.} \bibnamefont{Helstrom}},
  \emph{\bibinfo{title}{Quantum Detection and Estimation Theory}}
  (\bibinfo{publisher}{Academic Press}, \bibinfo{address}{New York},
  \bibinfo{year}{1976}), chap.~\bibinfo{chapter}{IV}.

\end{thebibliography}
\end{document}